\documentclass[%
 reprint,
 amsmath,amssymb,
 aps, prl
]{revtex4-2}
\usepackage{graphicx}
\usepackage{textcomp}
\usepackage{amssymb}
\usepackage[colorlinks,linkcolor=blue,anchorcolor=blue,citecolor=blue,urlcolor=black]{hyperref}
\usepackage[utf8]{inputenc}
\usepackage{amsmath}
\usepackage{tipa}
\usepackage{comment}
\usepackage{mathtools}
\usepackage{braket}
\usepackage{listings}
\usepackage{xcolor}
\usepackage{soul}
\usepackage{etoolbox}
\usepackage{lineno}
\usepackage{ulem}
\usepackage{CJKutf8}
\usepackage{multirow}
\usepackage{ragged2e}
\appto{\bibfont}{\RaggedRight}

\definecolor{Darkgreen}{rgb}{0,0.4,0}
\definecolor{listinggray}{gray}{0.9}
\definecolor{lbcolor}{rgb}{0.9,0.9,0.9}
\begin{document}

\title{Fundamental impossibility of a superradiant neutrino laser} 

\author{Yu-Kun Lu}
\affiliation{Department of Physics, Massachusetts Institute of Technology, Cambridge, MA 02139, USA}
\affiliation{Research Laboratory of Electronics, Massachusetts Institute of Technology, Cambridge, MA 02139, USA}
\affiliation{MIT-Harvard Center for Ultracold Atoms, Cambridge, MA, USA}

\author{Hanzhen Lin (\begin{CJK*}{UTF8}{gbsn}林翰桢\end{CJK*})}
\affiliation{Department of Physics, Massachusetts Institute of Technology, Cambridge, MA 02139, USA}
\affiliation{Research Laboratory of Electronics, Massachusetts Institute of Technology, Cambridge, MA 02139, USA}
\affiliation{MIT-Harvard Center for Ultracold Atoms, Cambridge, MA, USA}

\author{Wolfgang Ketterle}
\affiliation{Department of Physics, Massachusetts Institute of Technology, Cambridge, MA 02139, USA}
\affiliation{Research Laboratory of Electronics, Massachusetts Institute of Technology, Cambridge, MA 02139, USA}
\affiliation{MIT-Harvard Center for Ultracold Atoms, Cambridge, MA, USA}

\begin{abstract}

Here we address the fundamental question of whether an idealized system of $N$ atoms will show collective behavior and superradiance when it emits fermions instead of photons. We show that for single-fermion emission processes, the maximum emission is $\propto N$ and not $\propto N^2$, which proves the absence of superradiance and shows that the recent proposal to realize a superradiant neutrino laser is impossible. This can be understood as either destructive interference of fermionic transition amplitudes, or Pauli blockade by collective excitations with fermionic nature. We derive the exact solution of the fermionic Dicke problem and analyze the decay dynamics in various regimes. We extend the proof to arbitrary Hamiltonians and show that the jump rate operator for neutrino emission has a maximum eigenvalue of $N$ times the single-particle rate $\Gamma_0$. States with low excitation can show collective behavior and emit at a rate of $N \Gamma_0$.

\end{abstract}
\maketitle
\textbf{Superradiance and statistics. }
Collective emission of light by atoms leads to superradiance, a coherence-brightened laser~\cite{DickePhysRev.93.99,Gross_Haroche_review_1982} which works without a cavity. $N$ atoms emit light with a rate proportional to $N^2$ due to constructive interference of the amplitudes for spontaneous emission. This paradigmatic system has been discussed in various idealized models and realized in different systems~\cite{Gross_Haroche_review_1982,Book_SR, PhysRevA.40.4467,ketterle2001cargese,scheibner2007superradiance}.

In the original papers by Dicke, the emitters are localized, and their quantum statistics do not play any role~\cite{Dicke1964}. This is different for delocalized atoms, e.g., in a Bose-Einstein condensate (BEC). After the experimental observation of Rayleigh superradiance in a BEC, there was a controversy whether this relies on coherent matter wave amplification and would therefore be impossible in thermal or fermionic clouds. It was predicted that such superradiance is possible, despite additional Doppler broadening and a shortening of the coherence time caused by thermal or Fermi velocity~\cite{KetterleMWA,Meystre_PhysRevLett.86.4199}, and this was experimentally confirmed~\cite{PhysRevLett.94.083602,Fermion_SR_PhysRevLett.106.210401}. Of course, if the momentum of the emitted photon is smaller than the Fermi momentum, there
is Pauli blocking~\cite{Pauli_Blocking,Sanner,Deb}. Rayleigh superradiance in atomic gases can be understood as the amplification of a bosonic excitation (density modulation or phonon)~\cite{KetterleMWA}, which is possible in both fermionic and bosonic gases. Numerous papers explored various aspects of superradiance in fermionic systems~\cite{PhysRevLett.112.143002,PhysRevLett.112.143003,PhysRevLett.112.143004,SR_and_Quasicrystal_PhysRevResearch.7.013056,SR_Phase_transition_FeFermi_gas}.

However, almost all discussions of superradiance and quantum statistics focused on the case of fermionic emitter. Here we raise the question: What happens if the emitted particle is a fermion? In principle, there are two-level systems which emit neutrinos via electron capture, or one could speculate about collective emission of photinos, the fermionic supersymmetric partner of the photon. Some aspects of emission of fermionic particles have been discussed in the context of the dissociation of a bosonic molecule into two fermionic constituents~\cite{han_pu_molecular_dissociation_2005,Kherun_PhysRevLett.96.110401}, or for the case of localized fermionic impurities coupled to the continuum~\cite{eugene_demler_fermionic_defect_matter_wave_2024} as a model for fermionic quantum optics. Other aspects of fermionic superradiance are discussed in the context of black hole physics and Hawking radiation~\cite{BlackHoleSR}.

Our work was motivated by a recent proposal to use a radioactive BEC of $^{83}\mathrm{Rb}$ atoms which decays to $^{83}\mathrm{Kr}$ via electron capture to realize a superradiant neutrino laser~\cite{neutrino_laser_jones_formaggio}. It was predicted that superradiance can reduce the decay lifetime from 86 days ($10^7$ s) to 1 min, an increase of the decay rate by 5 orders of magnitude, through collective decay of $N=10^6$ atoms.
In an accompanying paper, we analyze in general whether it is possible to use a BEC to enhance radioactive decay at MeV energies and conclude that the recent proposals~\cite{MARMUGI2018,neutrino_laser_jones_formaggio} neglect the multi-mode aspect and the short coherence time (due to recoil energies and subsequent decay) and have gains of only $10^{-17}$or less~\cite{lin2025can}. This applies to all proposals independent of the specific decay and symmetry of the emitted particle.

Here we focus on the question of fermionic symmetry of the emitted particles in a system where all limitations of multi-mode emission and recoil are absent. We assume an idealized system with an ensemble of atoms (either bosonic or fermionic) which decay by neutrino emission to a daughter atom of the opposite statistics. We assume that the atoms are individually localized to a region much smaller than the neutrino wavelength (e.g. by assuming emission of neutrinos at sub-eV energies) to eliminate recoil energies and Doppler shifts. We also assume a stable daughter atom without a K shell vacancy (which would decay on a fs scale). This could be realized with a highly ionized parent with only one electron, e.g. $^{83}\mathrm{Rb}^{36+}$. The daughter ion (e.g. $^{83}\mathrm{Kr}^{36+}$) is electronically stable since there are no electrons left.

\begin{figure}[h]
    \centering
    \includegraphics[width= \linewidth]{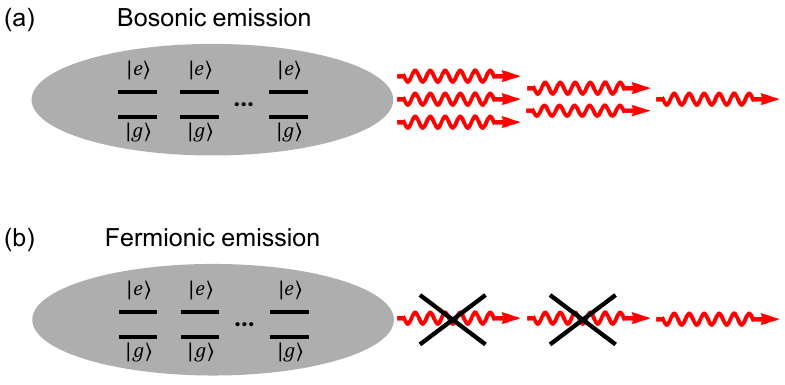}
    \caption{
    Superradiant ``laserlike'' emission of light has been observed from sodium BEC after laser excitation~\cite{SR_Rayleigh}. Recently, it has been suggested that radioactive BEC can be used to realize neutrino lasers~\cite{neutrino_laser_jones_formaggio}. However, as we explain in this paper, fermionic emission cannot be enhanced.
    }
    \label{fig:illustration}
\end{figure}

These assumptions lead to the ideal Dicke model~\cite{DickePhysRev.93.99, Dicke1964} formulated for the emission of neutrinos instead of photons. Our main result is the analytic solution of this model including classification of all states. We show that superradiance in such a system is absent --- the maximum possible rate of spontaneous emission is $N \Gamma_0$ in contrast to photons, where it is $N(N+1) \Gamma_0/4$, where $\Gamma_0$ is the natural decay rate of a single atom.

In the neutrino laser proposal, the authors argued that Pauli blocking should not play any role if the neutrinos are emitted sufficiently slowly so there is $\ll 1$ neutrino in the system at any given time~\cite{neutrino_laser_jones_formaggio}. There is also no Pauli blocking for the fermionic parent or daughter atom when the atoms are individually localized. However, our analysis shows that fermionic emission creates collective fermionic excitations, and Pauli blocking of these excitations restricts the whole system to emit only one fermion in the absence of state mixing by decoherence (Fig.~\ref{fig:illustration}). Furthermore, we extend the neutrino Dicke model to arbitrary many-body systems, analyze multi-mode emission and introduce additional forms of dissipation which we discuss with Lindblad master equations.

\textbf{Description of the fermionic Dicke system}. 
We consider a system of $N$ atoms tightly localized in individual traps, and the whole system is much smaller than the wavelength $\lambda$ of the emission. We consider the case that the parent atom is a boson and the daughter a fermion, but the behavior is almost identical in the opposite case (see End Matter). Similar to many discussions of photonic superradiance, we assume the atoms to interact only by coupling to the same mode of the neutrino field and neglect other interactions between the atoms, analogous to neglecting dipole-dipole interactions in the photonic case~\cite{Gross_Haroche_review_1982,footnote1}. 
We use $b$, $f$, $\nu$ as the annihilation operators for the emitting parent, daughter and emitted neutrino, state $\ket{e}$ and $\ket{g}$ to represent a site occupied by the parent and daughter, respectively. The decay can be described by the operator $ \nu^{\dagger}f_i^{\dagger} b_i$ on site $i$ ($0\leq i\leq N-1$).
As the neutrino leaves the system quickly and can be adiabatically eliminated (see End Matter), we can describe the effect of neutrino emissions by the on-site jump operators $c_i \equiv f_i^{\dagger} b_i$ which annihilate an excitation of the system by flipping one atom from the bosonic parent to the fermionic daughter state. We can verify that $c_i$ satisfy the anti-commutation relation for fermion operators: $\left\{c_i, c_j\right\}=0$ and $
\left\{c_i, c_j^{\dagger}\right\} =\left(f_i^\dagger f_i+b_i^\dagger b_i\right)\delta_{ij} = \delta_{ij}$. In the last equality, we used the fact that there is always one atom per site. All atoms couple identically to the neutrino mode (since the system size $\ll \lambda$), so the neutrino emission occurs via the collective jump operator $L =\sqrt{N\Gamma_0}C$ with $C\equiv \sum_i c_i/\sqrt{N}$. For photonic emission, the definitions are analogous, but since the parent and daughter have the same statistics, $c_i$ on different sites commute, and $L$ is proportional to the lowering operator $S^-$ for the collective spin, which (via the Holstein–Primakoff transformation) approximates a bosonic annihilation operator. Note that for neutrino emission, the jump operator is \textit{not only} the isospin lowering operator $\tau^-$ as assumed in~\cite{neutrino_laser_jones_formaggio}, which would be analogous to the bosonic $S^-$ operator for photons. The correct jump operator is the product of the nuclear part $\tau^-$ with the lepton part (which is an electron annihilation operator), and is fermionic.

\textbf{No coherent addition of fermionic amplitudes}. 
In superradiance of photons, the emission rate can be proportional to $N^2$ due to constructive interference of the transition amplitudes.
Now we show this is fundamentally impossible for neutrino emission. Consider having $N$ atoms where each is formally in a superposition between parent ($\ket{e}$) and daughter ($\ket{g}$) state $\ket{\Psi}=2^{-N/2}\left(\ket{g}+\ket{e}\right)^{\otimes N}$~\cite{footnote2}. 
Using the collective jump operator, the total emission rate is given by $R=\bra{\Psi}L^\dagger L\ket{\Psi} = \Gamma_0 \sum_{i,j}\bra{\Psi}c_i^{\dagger}c_j\ket{\Psi}$. In the bosonic case, since the jump operators $c_i$ commute, one finds $\bra{\Psi}c_i^{\dagger}c_j\ket{\Psi}=1/4+\delta_{ij}/4$ and $R=\Gamma_0N(N+1)/4$.

For fermionic emission, the jump operators for different sites \textit{anticommute}, so we have $\bra{\Psi}c_i^{\dagger}c_j\ket{\Psi}=\delta_{ij}/2-(\delta_{i,j-1}+\delta_{i-1,j})/4$ and $R=\Gamma_0/2$ (see End Matter). One can ask if the cancellation of matrix elements for neutrino emission can be compensated for by preparing states on site $i$ with different phases $\left(\ket{g}+ e^{i\alpha_i}\ket{e}\right)/\sqrt{2}$. The total emission rate depends only on phases difference $\phi$ between neighboring sites, and one obtains $R = \Gamma_0(\frac{N-1}{2}(1-\cos(\phi))+\frac{1}{2})$ with an upper bound of $R=\Gamma_0(N-1/2)$ for $\phi = \pi$. Therefore, the collective decay rate of fermionic emission is bounded by $N\Gamma_0$, similar to incoherent radiation, in contrast to the bosonic result $R \approx N^2\Gamma_0/4$. This demonstrates that emission amplitudes do not add constructively for fermions and implies the absence of neutrino superradiance. 

\begin{figure}
    \centering
    \includegraphics[width= \linewidth]{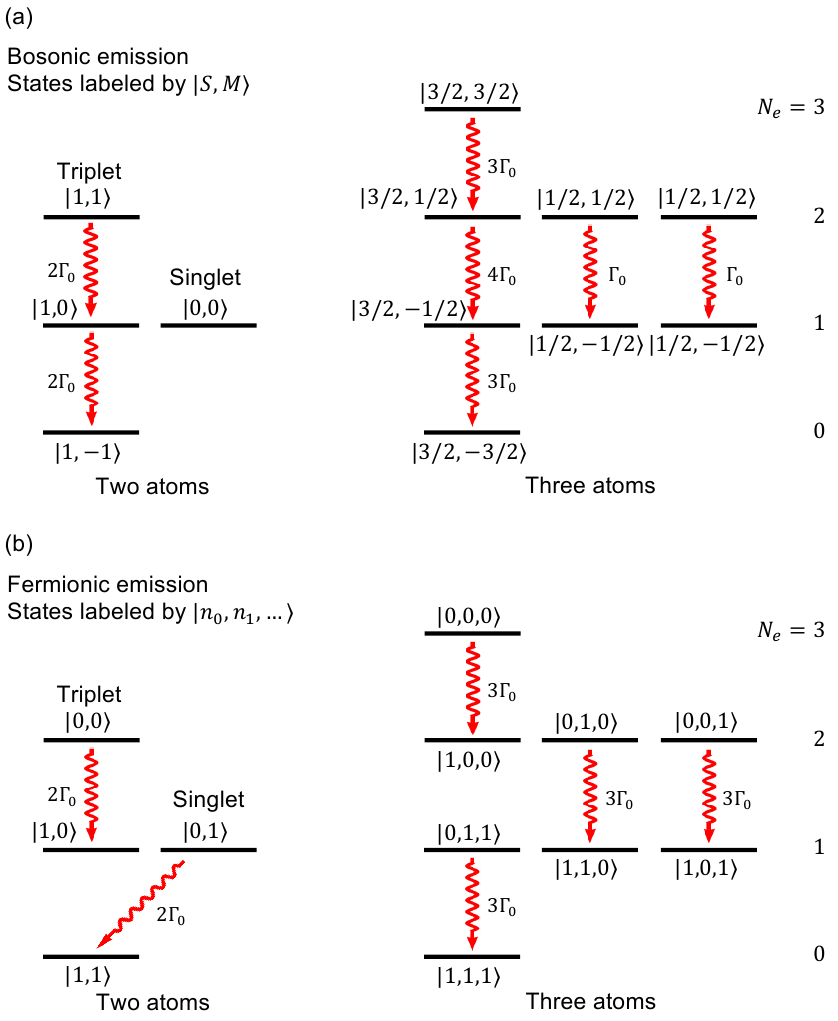}
    \caption{
    Comparison of the collective emission process for bosonic and fermionic radiation. In the bosonic case, the emission occurs along Dicke ``ladders'', whereas for fermions the emission stops after emitting a single neutrino due to Pauli blocking by a fermionic excitation left behind after the emission. For neutrino emission, the states form coupled pairs of bright and dark states.
   }
    \label{fig:levels}
\end{figure}

\textbf{No emission cascade for fermionic radiation.}
Since the fermionic jump operators $c_i$ obey fermionic anticommutators, it follows immediately that $L^2=0$ for the collective jump operator. Therefore, an arbitrary state can emit at most one neutrino and enter a dark state, where neutrino emission cannot occur due to destructive interference. 
Therefore, the radiation properties of the $2^N$ states of the system are very different from the bosonic case. We explicitly show the solutions for $N=2$ and $N=3$ (Fig.~\ref{fig:levels}). In the bosonic case, the states form vertically connected ladders, each corresponding to a specific value of the total spin $S$. The collective $S^-$ jump operator causes sequential transitions from the highest to the lowest state for a given $S$. In contrast, the fermionic emission always ends after one emitted neutrino. For $N=2$, the triplet state with one excitation is dark, whereas the singlet state $(\ket{ge} -\ket{eg})/\sqrt{2}$ decays to the triplet ground state, i.e. fermionic emission can connect an antisymmetric to a symmetric state. The analytic solution for three particles illustrates similar features: the energy level diagram breaks up into pairs of a bright state and a dark state, connected by the same transition rate $3 \Gamma_0$. The bosonic case shows the highest transition rate for the fully symmetric states with $S=3/2$ and a speedup between the first and second photon emission (Fig.~\ref{fig:levels}). 

To group and label the states connected by decay for $N$ atoms (Fig.~\ref{fig:levels}), we solve the eigenvalue problem for the emission rate operator $L^\dagger L$ and the excitation number $N_e=\sum b_i^\dagger b_i$. There are only two eigenvalues for $L^\dagger L$, $N\Gamma_0$ and 0, which correspond to the bright and dark states. Each bright state is paired with a dark state by the jump operator $L$. The labeling of degenerate states with the same $N_e$ is defined up to a unitary transformation.

This procedure can be formalized by defining operators for collective excitations $\tilde{f}_k^\dagger=\frac{1}{\sqrt{N}} \sum_j f_j^{\dagger}b_j e^{i j \frac{2 \pi k}{N}}, 0\leq k\leq N-1$. $\tilde{f}_k$ also fulfills the commutation relations for fermions. So instead of using localized fermionic daughter atoms basis $f_i^\dagger$, we transform to collective fermionic excitations basis $\tilde{f}_k^\dagger$. The important mode is $k=0$ since the collective jump operator $L= \sqrt{N\Gamma_0} C=\sqrt{N\Gamma_0} \tilde{f}_0^\dagger$. All other collective modes can be arbitrarily chosen to be orthogonal to each other and to $\tilde{f}_{k=0}$. 

By classifying the states using occupation numbers in the collective $\tilde{f}_k$ modes, we can single out the states with an unpopulated $k=0$ mode: they can emit exactly one neutrino since the emission can create an excitation in this mode. This excitation will now Pauli block further emission of neutrinos. This structure fully explains why a system of bosonic atoms can emit at most one neutrino, despite the apparent absence of direct Pauli blocking via the neutrinos or separately localized daughters. The Pauli blocking occurs through the fermionic collective excitation $\tilde{f}_{k=0}^\dagger$.
The initial state of all atoms in the bosonic parent state is the vacuum of all collective fermionic modes $\ket{00\cdots 0}$ where the labels are the occupation numbers of the $N$  modes. The initial state decays to the state $\ket{10\cdots 0}$. The lowest state is the state $\ket{11\cdots 1}$ and is connected to the bright state $\ket{01\cdots 1}$. All transition rates out of bright states have the same value $N \Gamma_0$.

For an arbitrary initial state $\ket{\psi}$, the emission rate is $R=N\Gamma_0\langle\psi|C^\dagger C|\psi\rangle=N\Gamma_0 n_0 \leqslant N\Gamma_0$, with $n_0= \langle C^\dagger C\rangle$ being the bright mode population. Therefore, $N\Gamma_0$ is the fastest collective decay rate possible.

We note that Dicke superradiance with localized emitters does not require identical atoms. They can be distinguishable, but they must couple to the radiation field in the same way. The same applies to the fermionic Dicke model. This emphasizes that the impossibility of neutrino superradiance is not necessarily related to Pauli blocking by fermionic daughter atoms, it is caused by Pauli blocking of fermionic excitations in the radiating system.

Note that our proof for the absence of neutrino superradiance is also valid when the $N$ atoms are part of a many-body system, or are bound into molecules with each other or another species. In the simple Dicke system, the Hamiltonian and the jump rate operator commute and we could find simultaneous eigenstates.
For an arbitrary many-body system, the collective jump operator for neutrino emission with wavevector $k$ is:
\[
L_{\mathbf{k}}
=
\int d^3\mathbf{r}\,
e^{i\mathbf{k}\cdot \mathbf{r}}\,
\psi_f^\dagger(\mathbf{r})\psi_b(\mathbf{r}),
\]
where $\psi$ are the field operators in second quantization.

The expression for the anti-commutator
\[
\{L_{\mathbf k},L_{\mathbf k}^\dagger\}
=
\int d^3r\,
(\psi_b^\dagger(\mathbf r)\psi_b(\mathbf r)+
\psi_f^\dagger(\mathbf r)\psi_f(\mathbf r))
=
\hat N_b+\hat N_f.
\]
 implies that
in the subspace of $N=\braket{\hat N_b+\hat N_f} $ particles in states $b$ and $f$, the normalized jump operator $c_k=L_{\mathbf k}/\sqrt{N}$ is a fermionic creation operator with $\{c_{\mathbf k},c_{\mathbf k}^\dagger\}=1$ and $c_{\mathbf k}^2=0$. The normalized jump rate operator $c_{\mathbf k}^\dagger c_{\mathbf k}$ is one minus the operator for fermionic occupation and has therefore eigenvalues of only 0 and 1, exactly the same as in the Dicke model. 
In case of a general neutrino mode with complex field $u(r)$, $N$ is replaced by the mode-intensity-weighted atom number $N_\mathrm{eff}=\int d^3r\, |u(\mathbf r)|^2
\left[
\psi_b^\dagger(\mathbf r)\psi_b(\mathbf r)
+
\psi_f^\dagger(\mathbf r)\psi_f(\mathbf r)
\right]
$. The jump rate is now bounded by 
$\langle L_{\mathbf k}^\dagger L_{\mathbf k} \rangle = N_\mathrm{eff} - \langle L_{\mathbf k} L_{\mathbf k}^\dagger \rangle \leq N_\mathrm{eff}
$.
The result is the same for fermionic parent atoms decaying into bosonic daughter atoms.

\textbf{Dynamics.} 
We now present various solutions for the decay dynamics of the Dicke system. For this, we describe the coherent coupling between the atoms and neutrinos with the Hamiltonian 
$H= g \sum_i \nu^{\dagger}f_i^{\dagger} b_i + \mathrm{h.c.}=g\sqrt{N}\left(\nu^\dagger C + C^\dagger\nu \right)$~\cite{ketterle2001cargese,Kherun_PhysRevLett.96.110401}. Here, $g$ represents the coupling strength for the decay process of the bosonic parent atom into the fermionic daughter atom and the neutrino. This Hamiltonian assumes that the neutrino is emitted into a single mode of a ``neutrino cavity''. 

This Hamiltonian describes a unitary time evolution, where a single neutrino is emitted and reabsorbed by the atoms, leading to Rabi oscillations in the bright mode population $n_0(t)=\cos^2(g\sqrt{N}t)$. 

Note that the Heisenberg equations of motion for fermionic emission differ by a minus sign compared to the case of bosons~\cite{Kherun_PhysRevLett.96.110401}. As a result, at early time, or with the undepleted source approximation,
one obtains exponential cosh and sinh functions for the bosonic case instead of sinusoidal Rabi oscillations for the fermionic case~\cite{Kherun_PhysRevLett.96.110401}.

We can now add a damping term $\kappa$ to the neutrino cavity. For the weakly damping case $\kappa<g\sqrt{N}$, the population in the bright mode exhibits damped oscillations with $ n_0(t)\approx{e^{-\kappa t/2}}\left(1+\cos(2g\sqrt{N}t)+\frac{\kappa \sin(2g\sqrt{N}t)}{4g\sqrt{N}}\right)/2$.
In the strong loss limit $\kappa\gg g\sqrt{N}$, the neutrino mode can be adiabatically eliminated, and the bright mode population decays exponentially $n_0(t)=\exp{\left(-4Ng^2t/\kappa \right)}=\exp{\left(-N\Gamma_0 t\right)}$ with $\Gamma_0=4g^2/\kappa$ being the single-particle decay rate~\cite{eugene_demler_fermionic_defect_matter_wave_2024}. 
The same derivation for photons leads to an initial exponential gain for bosonic emission which is the usual superradiance.

The discussions above assumed no dephasing mechanism for the atomic states, so only one neutrino emission is possible. In reality, effects such as Doppler effects introduce inhomogeneous broadening mechanisms and lead to dephasing with rate $\kappa_\phi$. The dephasing transfers population of dark modes into bright modes, and enables subsequent neutrino emission after dephasing time $1/\kappa_\phi$. 
In the weak dephasing limit $\kappa_\phi\ll N\Gamma_0$, each step of neutrino emission occurs at rate $N\Gamma_0$, but there needs to be a dephasing time $1/\kappa_\phi$ before the next emission occurs.
In the strong dephasing limit $\kappa_\phi\gg N\Gamma_0$, rapid dephasing occurs within $1/\kappa_\phi$. The collective nature of the decay disappears, and each particle decays with the single-particle decay rate $\Gamma_0$. This result emphasizes that the dark states are collective excitations that require phase-coherence between the atoms. Dephasing eliminates the Pauli blocking by the population of the collective mode $\tilde{c}_0$.
The solution of the full dynamics can be found in End Matter.

\begin{figure}
    \centering
    \includegraphics[width= \linewidth]{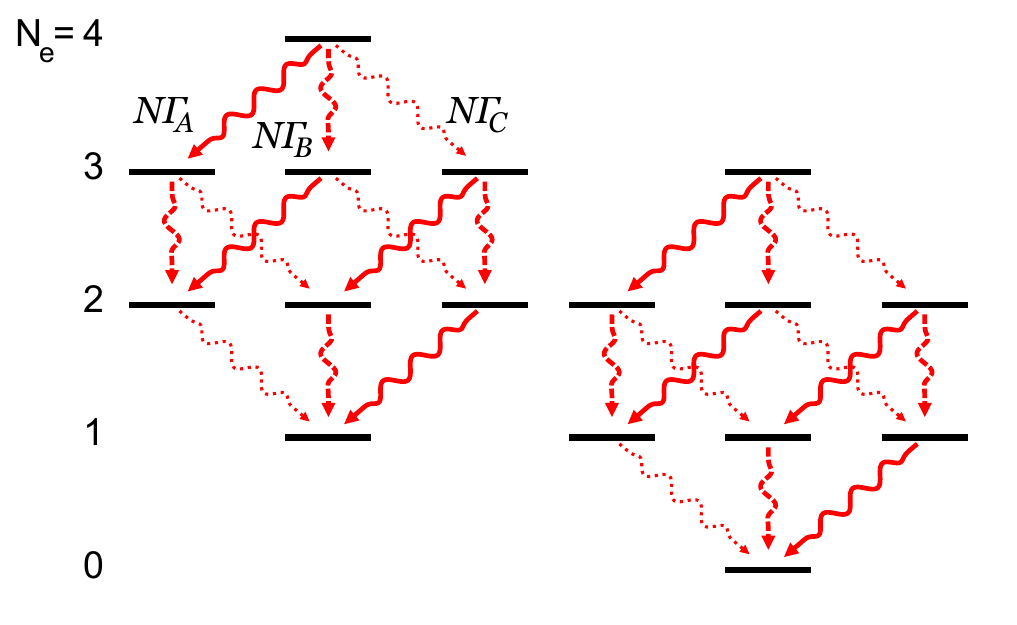}
    \caption{Fermion emission structure for 4 atoms and 3 modes. The 16 total possible states are grouped into two sectors of 8 states each.  Each sector is coupled by solid, dashed and dotted lines representing fermion emission into mode A, B and C. Connectivity in each sector is identical to a cube. For two modes, the dotted lines are eliminated, and the states fragment into 4 groups of 4 states.}
    \label{fig:multimode_emission_structure}
\end{figure}

 \textbf{Extended Samples.} In an extended sample with size $\ge\lambda$, the $k=0$ mode is not the only mode for neutrino emission, and the problem becomes multi-mode. We can easily extend our model to two and three orthogonal modes. Two modes could be realized with an array of emitters in a 1D waveguide~\cite{eugene_demler_fermionic_defect_matter_wave_2024}. 
 Using the same eigenstates as before, each state is now connected to two (three) other states through absorption or emission. A maximally bright state can now emit two (three) neutrinos and has then reached a dark state. For two allowed modes, the Hilbert space is segmented into groups of four states in the form of a rhombus (Fig.~\ref{fig:multimode_emission_structure}, solid and dashed lines). The emission rates are now 0, $N\Gamma_A$, $N\Gamma_B$, and $N(\Gamma_A + \Gamma_B)$.
 For three allowed modes the Hilbert space fragments into sectors of eight states with the same connectivity as a cube (Fig.~\ref{fig:multimode_emission_structure}, all lines).
If $n$ modes are allowed, the Hilbert space is separated into sectors each with $2^n$ states connected as a $n$-dimensional hypercube. The maximum possible emission rate is $N\Gamma_0$ where now $\Gamma_0 = \sum_{i=1}^n \Gamma_i$. The $N$ scaling implies absence of superradiance. Eventually, when the number of modes is comparable to the number of atoms,  the system will no longer show any collective effects and decay exponentially~\cite{lin2025can}.

\textbf{Pauli blocking by neutrinos.}
Ref.\,\cite{neutrino_laser_jones_formaggio} claimed that superradiance should be independent of the quantum statistics of the radiation as long as much less than one neutrino (or photon) would be present in the system. This condition could be physically imposed by surrounding the emitting condensate with an absorber. However, since this is equivalent to have them propagate into empty space, this creates a contradiction. For superradiance of photons, $N$ photons are emitted in a single pulse of duration $\approx 1/G$ with a Fourier-limited frequency spread. Here $G=N \Gamma \Omega$ is the emission rate into a superradiant mode of solid angle $\Omega$~\cite{lin2025can}. The temporal coherence of the superradiant pulse also follows from the semiclassical picture that the superradiant pulse is emitted by a macroscopic collective dipole moment (or Bloch vector) precessing at a constant frequency (the transition frequency of the two-level system). Therefore, all $N$ photons populate a single temporal mode in the form of the uncertainty-limited wave packet. Similarly, a hypothetical superradiant neutrino pulse of duration $\tau$ would put all the neutrinos into a single mode of extension $c \tau$, clearly a contradiction. This implies that neutrino superradiance is impossible due to Pauli blocking of neutrinos. This argument also implies that in the case of neutrinos, it is impossible to have a macroscopic Bloch vector, i.e., transition matrix elements of many atoms which add up constructively.

We want to point out that stimulation by atomic coherence (i.e. superradiance) and stimulation by the radiation (i.e. lasing) are not separable. There are two different ways to understand photon superradiance in extended ensembles~\cite{Gross_Haroche_review_1982,Schneble2003}. One can regard the whole system as a radiating system, and every photon which comes out is the result of spontaneous emission into empty space. One can also use a local description, where a photon is emitted at one point and travels through the system. This photon is necessarily amplified since the threshold for superradiance requires the optical density to be larger than one. 
Experimentally, it was observed that more recoiling atoms were produced at the ends of the elongated condensate, clear evidence for light propagating and being amplified within the condensate~\cite{Schneble2003}. Since the transition rate for fermionic fields is suppressed by $1-n$ instead of the bosonic enhancement factor $1+n$ (where $n$ is the occupation of the relevant mode), this is impossible for neutrinos. Interestingly, our solution of the fermionic Dicke problem adiabatically eliminated the emitted neutrinos, but then showed that only one neutrino can be emitted into a given mode without invoking the Pauli principle for neutrinos.

\textbf{Discussion.}
This work has shown the dramatic difference between fermionic and bosonic emission. In the bosonic case, the first emitted photon leaves behind a bosonic excitation (phonon, magnon etc.) --- this is, in the words of Dicke, the memory of the emitted photon~\cite{Dicke1964}, which enhances further excitation in the same mode (via bosonic stimulation). Fermionic emission is also imprinted into the memory of the emitting system, but via a collective fermionic excitation that blocks any emission into the same mode, as illustrated in figure~\ref{fig:levels}. The dephasing model discussed above introduces ``loss of memory'', and then the system can emit again.

In conclusion, we have proven that fermionic superradiance (i.e. $N^2$ scaling of the maximum emission rate) is impossible when neutrinos are emitted. Our proof applies to all systems where nuclei emit single neutrinos since the collective emission operator is fermionic and the jump rate operator has a maximum eigenvalue of $N \Gamma_0$.
In contrast, for emission of neutrino pairs (e.g. two-neutrino double electron capture~\cite{two_neutrino_double_electron_capture_124Xe})
the jump operator is bosonic, and in analogy to two-photon superradiance, enhanced decay could be possible~\cite{Two_photon}. However, the half-lives for those decays is longer than $10^{22}$ years~\cite{two_neutrino_double_electron_capture_124Xe}, a trillion times the age of the universe.

Furthermore, we have shown that some form of collective emission of fermions is possible. All bright states in the fermionic Dicke ladder have a rate of emission of $N \Gamma_0$ even if the number of excitations (bosonic parent atoms) $N_e \ll N$. This includes the state with $N_e=1$: the first fermionic Dicke state has the same enhanced emission rate $N \Gamma_0$ as the first bosonic Dicke state. This is almost trivial, since the $N$-fold enhancement is the time reversal of the fact that $N$ ground state atoms absorb a single photon or neutrino with a probability proportional to $N$, but it is still a collective enhancement. The first excited Dicke state can be regarded as the intermediate state in coherent elastic neutrino-nucleus scattering~\cite{coherent_neutrino_neucleus_scattering}. Collective effects could be experimentally studied in the dissociation of diatomic molecules into pairs of fermions~\cite{han_pu_molecular_dissociation_2005,Kherun_PhysRevLett.96.110401}. These are the same systems which feature the BEC-BCS crossover in their phase diagram~\cite{ketterle2008making}. Another platform is localized fermionic atoms in an optical lattice coupled via microwaves to propagating states~\cite{eugene_demler_fermionic_defect_matter_wave_2024}. So far, experiments have been done with bosonic atoms and explored superradiance and subradiance~\cite{Schneble2025}.
 

\textit{Acknowledgments}.---The authors are grateful to many people for helpful discussions, including Joe Formaggio, Ben Jones, James Thompson, Ana-Maria Rey, and our whole research group. We thank William Milner and Guoxian Su for comments on the manuscript. Our research has been supported by NSF (grant No. PHY-2208004), from the Center for Ultracold Atoms (an NSF Physics Frontiers Center, grant No. PHY-2317134), by the Vannevar-Bush Faculty Fellowship (grant No. N00014-23-1-2873), by the Gordon and Betty Moore Foundation (GBMF ID \# 12405), and by the Army Research Office (contract No. W911NF2410218). 

%

\newpage
\clearpage

\section{End Matter}

\newcommand{\cdag}{c^\dagger}
\newcommand{\nudag}{\nu^\dagger}
\newcommand{\nc}{n_C}
\newcommand{\nnu}{n_\nu}

\subsection{Fermions emitting fermions}
Almost identical results are obtained when the parent atom is fermionic. The collective jump operator for neutrino emission is now $L^\dagger$ instead of $L$. In the $N=2$ Dicke problem (Fig.~\ref{fig:levels}), the singlet state is now dark and the triplet state with $N_e=1$ is bright. For each atom in an equal superposition state between parent and daughter states, the total emission rate is now $\Gamma = \Gamma_0(\frac{N-1}{2}(1+\cos(\phi))+\frac{1}{2})$ and has its upper limit if all phase differences $\phi$ between neighboring sites are zero. If we define the collective mode operators via $\tilde{f}_k^\dagger=\frac{1}{\sqrt{N}} \sum_j f_jb_j^\dagger e^{i j \frac{2 \pi k}{N}}$, we arrive at the same picture that each neutrino emission would create population in the fermionic mode $\tilde{f}_0$. Using the alternative definition $\tilde{f}_k^\dagger=\frac{1}{\sqrt{N}} \sum_j f_j^\dagger b_j e^{i j \frac{2 \pi k}{N}}$ would result in the picture that the system initially is in a Fermi sea of collective excitations, and each neutrino emission creates a hole in the mode $\tilde{f}_0$. Now the emission stops after a single neutrino because we have Pauli blocking via a collective hole excitation.

\subsection{Calculation of the emission matrix elements}

Define the superposition state~\cite{footnote2}:
\begin{equation}
    |\Psi\rangle=\left(\frac{|g\rangle+|e\rangle}{\sqrt{2}}\right)^{\otimes  N}=\prod_{i=0}^{N-1} \frac{\left(f_i^{\dagger}+b_i^{\dagger}\right)}{\sqrt{2}}|0\rangle
\end{equation}

The emission rate is $\bra{\Psi} L^\dagger L\ket{\Psi}=\Gamma_0\sum_{i,j}\bra{\Psi} b_i^\dagger f_i f_j^\dagger b_j\ket{\Psi}$. To evaluate each matrix element, we push $f_j^\dagger b_j$ to the right side and $b_i^\dagger f_i$ to the left side:

\begin{widetext}
\begin{equation*}
\begin{aligned}
     &f_j^{\dagger} b_j \prod_{n=0}^{N-1}\left(\frac{f_{n}^{\dagger}+b_n^\dagger}{\sqrt{2}}\right)|0\rangle\\
     &=\prod_{n_1=0}^{j-1}\left(\frac{-f_{n_1}^{\dagger}+b_{n_1}^{\dagger}}{\sqrt{2}}\right)\left(\frac{-f_j^{\dagger}+b_j^{\dagger}}{\sqrt{2}} f_j^{\dagger} b_j+\frac{f_j^{\dagger}}{\sqrt{2}}\right) \prod_{n_2=j+1}^{N-1}\left(\frac{f_{n_2}^\dagger+b_{n_2}^{\dagger}}{\sqrt{2}}\right)\ket{0} \\
     &=\prod_{n=0}^{N-1}\left(\frac{-f_n^{\dagger}+b_n^{\dagger}}{\sqrt{2}}\right) f_j^{\dagger} b_j\ket{0}+\prod_{n_1=0}^{j-1}\left(\frac{-f_{n_1}^{\dagger}+b_{n_1}^{\dagger}}{\sqrt{2}}\right) \frac{f_j^{\dagger}}{\sqrt{2}} \prod_{n_2= j+1}^{N-1} \left(\frac{f_{n_2}^{\dagger}+b_{n_2}^{\dagger}}{\sqrt{2}}\right)\ket{0}\\
     &=\prod_{n_1=0}^{j-1}\left(\frac{-f_{n_1}^{\dagger}+b_{n_1}^{\dagger}}{\sqrt{2}}\right) \frac{f_j^{\dagger}}{\sqrt{2}} \prod_{n_2= j+1}^{N-1} \left(\frac{f_{n_2}^{\dagger}+b_{n_2}^{\dagger}}{\sqrt{2}}\right)\ket{0}.
\end{aligned}
\end{equation*}
\end{widetext}
Similar calculations can be done when pushing $b_i^\dagger f_i$ to the left. The matrix element is then obtained by taking the inner product of the two results.
If $i=j$, we have $\braket{b_i^\dagger f_i f_i^\dagger b_i}={1}/{2}$.
If $|i-j|=1$, we have contribution from indices $i$ and $j$ such that $\braket{b_i^\dagger f_i f_j^\dagger b_j}=-({1}/{2})^2=-1/4$.
If $|i-j|>1$, the matrix element $\quad\braket{b_i^\dagger f_i f_j^\dagger b_j}=0$ because of the contribution from $\left\langle{\left(-f_m+b_m\right)} ({f_m^\dagger+b_m^\dagger})\right\rangle=0$ for $i<m<j$.
Combine the results above, we obtain that $\left\langle c_i^{^\dagger} c_j\right\rangle=\frac{1}{2} \delta_{i j}-\frac{1}{4}\left(\delta_{i, j-1}+\delta_{i, j+1}\right)$. For the general case of $|\Psi\rangle=\left(\frac{|g\rangle+e^{i\alpha_i}|e\rangle}{\sqrt{2}}\right)^{\otimes  N}=\prod_{i=0}^{N-1} \frac{\left(f_i^{\dagger}+e^{i\alpha_i}b_i^{\dagger}\right)}{\sqrt{2}}|0\rangle$, we have $\left\langle c_i^{^\dagger} c_j\right\rangle=\frac{1}{2} \delta_{i j}-\frac{e^{i(\alpha_j-\alpha_i)}}{4}\left(\delta_{i, j-1}+\delta_{i, j+1}\right)$.

\subsection{Lindblad equation for atoms}

Here, we derive the time evolution of the entire system of parent and daughter atoms and neutrinos using Lindblad equations. We assume the neutrinos are emitted into a single mode of a ``neutrino cavity'' with decay rate $\kappa$. The Hamiltonian and jump operator read:
\begin{equation}
 H=g\sqrt{N}\big(\nudag C + C^\dagger\nu\big),\qquad L=\sqrt{\kappa}\nu\qquad.   
\end{equation}

The fermionic modes \(C,\nu\) obey
\begin{equation}
 \{C,C^\dagger\}=1,\quad \{\nu,\nudag\}=1,\quad
\{C,\nu\}=\{C,\nudag\}=0 .   
\end{equation}

The adjoint Lindblad equation for any operator \(O\) reads:
\begin{equation}
\dot O
= i[H,O]+\kappa\left(\nudag O \nu-\tfrac12\{\nudag\nu,O\}\right).    
\end{equation}

Define the (closed) set of second moments:
\begin{equation}
  \nc\equiv C^\dagger C,\qquad \nnu\equiv \nudag \nu,\qquad s\equiv \nu C^\dagger,\qquad s^\dagger\equiv C\nudag, 
\end{equation}

then we can write down their equation of motion. For convenience, we define \(s=r+iu\) with \(r=\text{Re}(s),\ u=\text{Im}(s)\). 
\begin{equation}
\begin{aligned}
\dot n_C  &= -2g\sqrt{N}u,\\
\dot n_\nu &= 2g\sqrt{N}u - \kappa\nnu,\\
\dot r    &= -\tfrac{\kappa}{2}r,\\
\dot u    &= g\sqrt{N}\big(\nc-\nnu\big) - \tfrac{\kappa}{2}u.
\end{aligned}    
\end{equation}

Now we can discuss the properties of the solutions in differnt regimes.
\paragraph{Steady state (with a non-zero $\kappa$):}
\begin{equation}
{\ \langle s\rangle_\infty=0,\quad \langle \nnu\rangle_\infty=0,\quad \langle \nc\rangle_\infty=0\ }.    
\end{equation}

This means that after a long time, the bright mode will be depleted by emitting one neutrino.

\paragraph{Lossless limit \((\kappa=0)\):}
\begin{equation}
 \frac{d^2}{dt^2}(\nc-\nnu)+4g^2N\big(\nc-\nnu\big)=0.   
\end{equation}

From initial condition \(\langle \nc(0)\rangle=1,\langle \nnu(0)\rangle=\langle u(0)\rangle=0\), we have the dynamics of populations:
\begin{equation}
    {\ \langle \nc(t)\rangle=\cos^2(g\sqrt{N}t),\langle \nnu(t)\rangle=\sin^2(g\sqrt{N}t)\ }.
\end{equation}
Physically, this means that the neutrino is emitted and reabsorbed by the atom, so the population of the bright mode and the neutrino mode exhibits Rabi oscillations.

\paragraph{Weak damping limit \((\kappa\ll g\sqrt{N})\):} In this regime, we have the physics of damped Rabi oscillation with 
\begin{equation}
    n_C(t)\approx\frac{e^{-\kappa t/2}}{2}\left(1+\cos(2g\sqrt{N}t)+\frac{\kappa \sin(2g\sqrt{N}t)}{4g\sqrt{N}}\right).
\end{equation}

\paragraph{Bad cavity limit \((\kappa\gg g\sqrt{N})\) (adiabatic elimination):}
\begin{equation}
 u \approx \frac{2g\sqrt{N}}{\kappa}\nc,\quad \nnu \approx \frac{4Ng^2}{\kappa^2}\nc\approx0,   
\end{equation}

\begin{equation}
    \frac{d}{dt}\langle \nc\rangle \approx -\frac{4g^2N}{\kappa}\langle \nc\rangle
\end{equation}

\begin{equation}
 \langle \nc(t)\rangle \approx \braket{\nc(0)}e^{-(4g^2N/\kappa)t}=\braket{\nc(0)}e^{-N\Gamma_0t}
\end{equation}

It shows that the population in the bright mode decays exponentially with rate $N\Gamma_0$ with $\Gamma_0=4g^2/\kappa$. This is analogous to the Purcell effect in cavity QED~\cite{purcell1995spontaneous}.

So far we have considered an idealized case without dephasing. Inhomogeneous broadening will introduce dephasing and modify the decay dynamics. We study this by adding the following jump operator to the Lindblad equation:

\begin{equation}
    L_{\phi,i}=\sqrt{\kappa_\phi}n_i,\qquad n_i\equiv c_i^\dagger c_i \quad (i=1,\dots,N).
\end{equation}

The adjoint Lindblad equation for any operator \(O\) now reads:

\begin{equation}
\begin{aligned}
    \dot O &= i[H,O]+\kappa\left(\nu^\dagger O\nu-\tfrac12\{\nu^\dagger\nu,O\}\right)\\
    &+ \kappa_\phi\sum_{i=1}^N\left(n_i O n_i-\tfrac12\{n_i,O\}\right). 
\end{aligned}
\end{equation}

Denoting the site average \(\displaystyle \bar n\equiv \frac{1}{N}\sum_i n_i\) and working out the commutators, one finds the closed set of equations:
\begin{equation}
  \begin{aligned}
\dot n_C &= -2g\sqrt{N}u-\kappa_\phi(n_C-\bar n),\\
\dot n_\nu&= 2g\sqrt{N}u-\kappa n_\nu,\\
\dot r &= -\tfrac{\kappa+\kappa_\phi}{2}r.\\
\dot u   &= g\sqrt{N}\big(n_C-n_\nu\big)-\tfrac{\kappa+\kappa_\phi}{2}u,\\
\dot{\bar n}&= -\frac{2g\sqrt{N}}{N}u,
\end{aligned} 
\end{equation}

The term \(-\kappa_\phi(n_C-\bar n)\) relaxes the bright-mode population towards \(\bar n\) by eliminating inter-site coherences; both \(\kappa\) and \(\kappa_\phi\) damp the pair coherence \(s\).

In the bad cavity limit ($\kappa\gg\kappa_\phi,g\sqrt{N}$), the fast variables \(u\) can be adiabatically eliminated:
\begin{equation}
 0\simeq \dot u = g\sqrt{N}(n_C-n_\nu)-\tfrac{\kappa}{2}u
\quad\Rightarrow\quad
u \simeq \frac{2g\sqrt{N}}{\kappa}n_C .  
\end{equation}

Putting this into the equations for \(\dot n_C,\dot{\bar n}\) gives
\begin{equation}
\begin{aligned}
\dot n_C &\simeq -N\Gamma_{0}n_C-\kappa_\phi(n_C-\bar n),\\[2pt]
\dot{\bar n}&\simeq -{\Gamma_{0}}n_C,
\end{aligned}    
\end{equation}

It is convenient to introduce the bright contrast $x\equiv n_C-\bar n$. In terms of \((\nc,x)\) the dynamics is linear:
\begin{equation}
  \frac{d}{dt}
\begin{pmatrix} n_C\\ x \end{pmatrix}
=
\begin{pmatrix}
-N\Gamma_{0} & \ \ -\kappa_\phi\\[3pt]
-N\Gamma_{0}(1-\tfrac1N) & \ \ -\kappa_\phi
\end{pmatrix}
\begin{pmatrix} n_C\\ x \end{pmatrix}.  
\end{equation}

The eigenvalues are
\begin{equation}
 \lambda_{\pm}= -\frac{N\Gamma_{0}+\kappa_\phi}{2}
\ \pm\ \frac12\sqrt{(N\Gamma_{0}+\kappa_\phi)^2-{4\Gamma_{0}\kappa_\phi}}\ <0,   
\end{equation}

so both modes decay exponentially, leading to the steady state
\begin{equation}
   \langle n_C\rangle_\infty=\langle \bar n\rangle_\infty=0,\quad \langle \nnu\rangle_\infty=0,\quad \langle s\rangle_\infty=0.
\end{equation}

For \(\kappa_\phi\ll N\Gamma_{\rm 0}\), the two decay rates are $\kappa_\phi/N$ and $N\Gamma_0$. This is the weak dephasing regime, where the first neutrino is emitted at a rate $N\Gamma_0$. But then the bright mode is depleted and the subsequent decay slows down. In other words, there is a wait time of $1/\kappa_\phi$ for dephasing to occur before the next neutrino can be emitted, so the time constant for the $N$ atoms to decay is $N/\kappa_\phi$.

For \(\kappa_\phi\gg N\Gamma_{\rm0}\), the two decay rates are $\kappa_\phi$ and $\Gamma_0$. Physically, this is the strong dephasing regime, and rapid dephasing happens within time $1/\kappa_\phi$. After that, the coherence between different atoms disappears, and each atom decays independently at the single-particle rate $\Gamma_0$.

\makeatother

\onecolumngrid

\end{document}